\documentclass[%
 reprint,
 amsmath,amssymb,
 aps,
]{revtex4-1}

\usepackage{graphicx}
\usepackage{dcolumn}
\usepackage{bm}
\usepackage{physics}
\usepackage{mathrsfs}
\usepackage{comment}
\usepackage{amsmath}
\usepackage[T1]{fontenc}

\begin{document}

\preprint{APS/123-QED}

\title{Broadband coherent optical memory based on electromagnetically induced transparency }

\author{Yan-Cheng Wei$^{1,2}$}
\author{Bo-Han Wu$^{1}$}
\author{Ya-Fen Hsiao$^{1}$}
\author{Pin-Ju Tsai$^{1,2}$}
\author{Ying-Cheng Chen$^{1,3}$}

\email{Corresponding author: chenyc@pub.iams.sinica.edu.tw}
\affiliation{$^{1}$Institute of Atomic and Molecular Sciences, Academia Sinica, Taipei 10617, Taiwan}
\affiliation{$^{2}$Department of Physics, National Taiwan University, Taipei 10617, Taiwan}
\affiliation{$^{3}$Center for Qauntum Technology, Hsinchu 30013, Taiwan}
\date{\today}

\begin{abstract}
Quantum memories, devices that can store and retrieve photonic quantum states on demand, are essential components for scalable quantum technologies. It is desirable to push the memory towards the broadband regime in order to increase the data rate. Here, we present a theoretical and experimental study on the broadband optical memory based on electromagnetically-induced-transparency (EIT) protocol. We first provide a theoretical analysis on the issues and requirements to achieve a broadband EIT memory. We then present our experimental efforts on EIT memory in cold atoms towards the broadband or short-pulse regime. A storage efficiency of $\sim 80 \%$ with a pulse duration of 30 ns (corresponding to a bandwidth of 14.7 MHz) is realized. Limited by the available intensity of the control beam, we could not conduct an optimal storage for the even shorter pulses but still obtain an efficiency of larger than $50 \%$ with a pulse duration of 14 ns (31.4 MHz). The achieved time-bandwidth-product at the efficiency of $50 \%$ is 1267.  

\end{abstract}

\maketitle

\newcommand{\FigOne}{
	\begin{figure}[t]
	{\includegraphics[width=7cm]{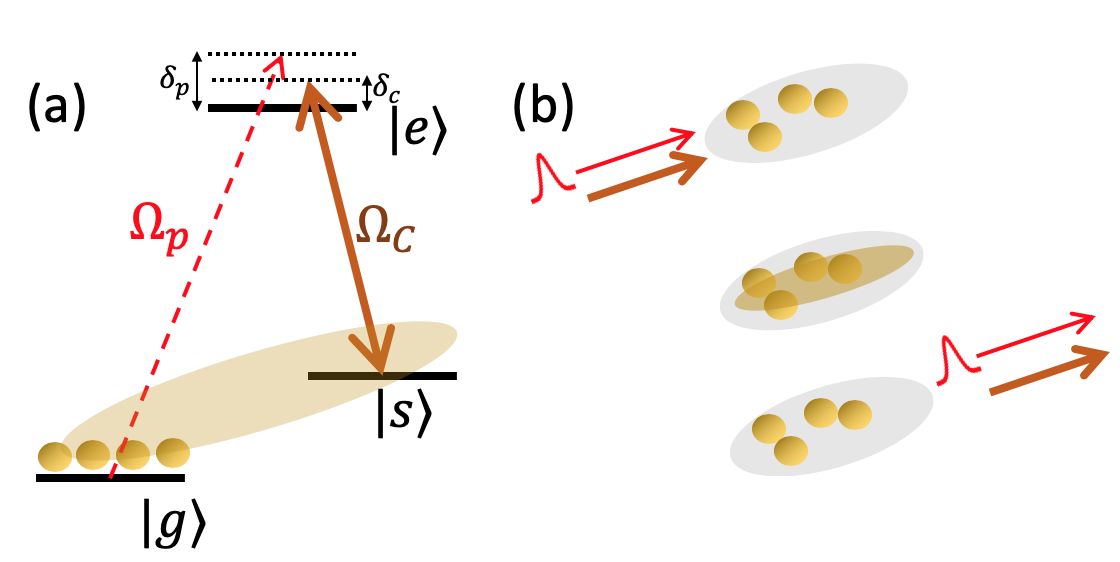}}
	\caption{ (a) Transition scheme of the $\Lambda$-type EIT system. $|g\rangle$ and $|s\rangle$ are the two ground states and $|e\rangle$ is the excited state. $\Omega_{p,c}$ stand for the Rabi frequency of the probe and control field, respectively. (b) Photonic storage through the EIT protocol. In both graphs, the yellow shaded area denotes the spin coherence in the two ground states.
	}
	\label{Fig1}
	\end{figure}
}
\newcommand{\FigTwo}{
	\begin{figure}[t]
	{\centering\includegraphics[width=8.5cm,viewport=80 0 890 450,clip]{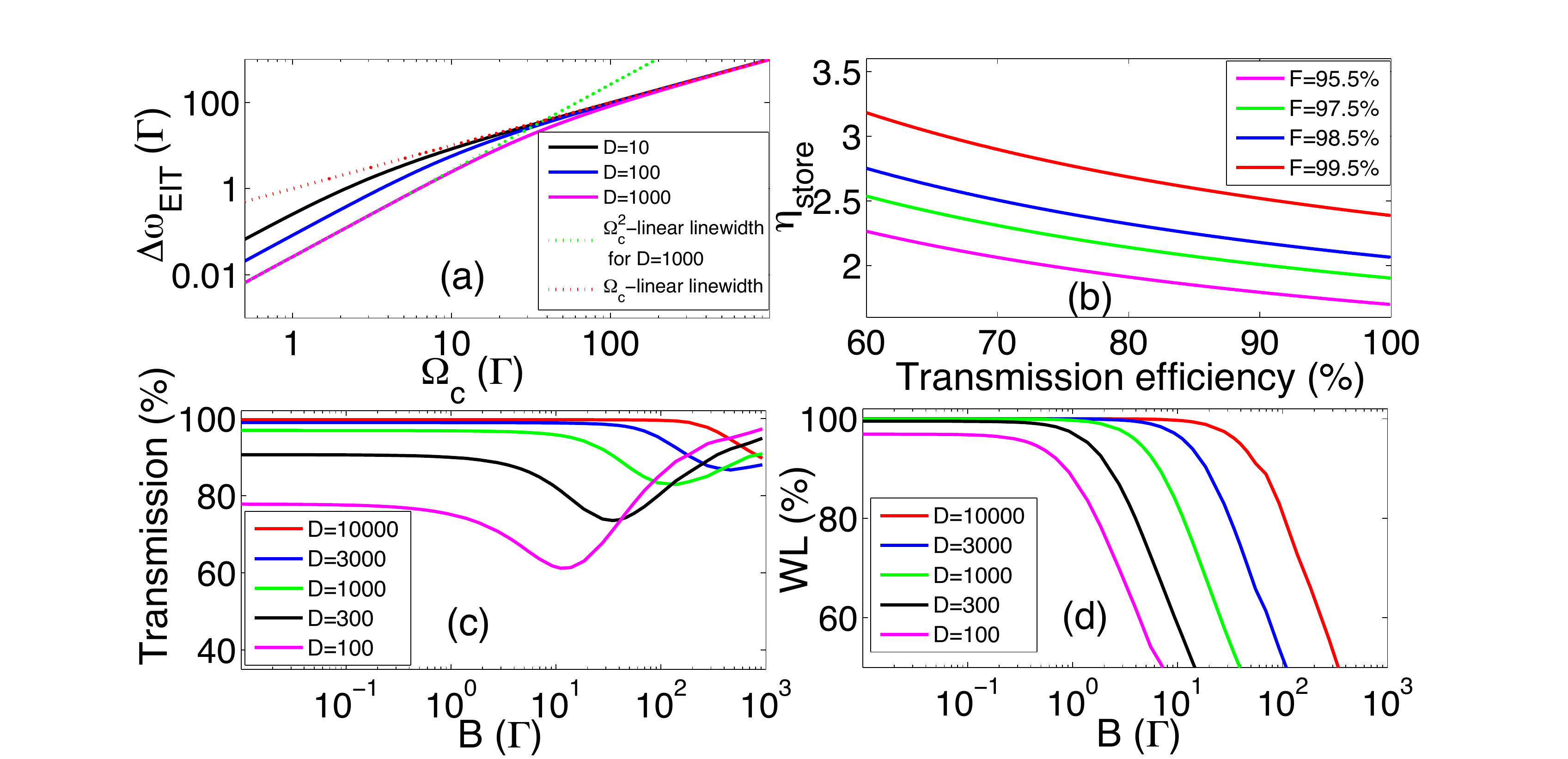}\\}
	\caption{ (a) Solid black, blue and pink curves denote the FWHM EIT transparent bandwidth ($\Delta \omega_{\text{EIT}}$) versus $\Omega_c$ for optical depth $D$ of 10, 100 and 1000, respectively. The green and red dotted line are the approximate formula of $\Delta \omega_{\text{EIT}}$ in the intensity-linear bandwidth regime for $D$=1000 and field-linear bandwidth regime, respectively. (b) The $\eta_{\text{store}}$ for the required TE is plotted. $F = 99.5\%, 98.5\%, 97.5\%, 95.5\%$ for the curves from top to bottom. (c) and (d) depict TE and WL versus different bandwidth of the input probe pulse ($B$) for various optical depths. Here, $\Omega_c$ for different $D$ are determined by $\eta=$2.43. 
	}
	\label{Fig2}
	\end{figure}
}




\nocite{*}

\section{Introduction}
Quantum memories are crucial components in linear-optics-based quantum computation and long-distance quantum communication based on quantum repeater protocol\cite{Bussires2013, Duan2001}. There have been significant efforts and progress on quantum memories in the past two decades. Parameters to evaluate the performance of a quantum memory include fidelity, efficiency, storage time, bandwidth or capacity, and noise level. It is still a great challenge to develop a quantum memory with a high performance on all aspects. In this work, we focus our discussion on achieving a broad bandwidth while maintaining a high efficiency for coherent optical memories based on EIT protocol.  

Optical memory based on the off-resonant Raman transition is generally considered to be advantageous in achieving a high bandwidth. Based on Raman memory protocol, there have been many works pushing the bandwidth towards $\sim$ 100 MHz-1 GHz range with an efficiency of ranging from 10-30 \%\cite{Reim2010,Reim2011,Sprague2014,Ding2015,Wolters2017,Finkelstein2018}. With the same protocol, one recent work achieved an efficiency of 82 \% at a bandwidth of $\sim$ 100 MHz\cite{Guo2019}. 

However, the on-resonant two-photon transition in a $\Lambda$-type atomic system also allows one to implement the broadband optical memory, as long as the optical depth of the media is high and the intensity of the control beam is strong enough\cite{Gorshkov2007,Liao2014,Saglamyurek2018,Rastogi2019}. The memory character is quite different depending on the ratio of the spectral bandwidth of the probe pulse (denoted as $B$) to the EIT transparent bandwidth ($\Delta\omega_{EIT}$). If $B$ is larger than $\Delta\omega_{EIT}$, the memory operation involves coherent absorption of the probe pulse by the two Autler-Townes absorption peaks and this is called the Autler-Townes-splitting (ATS) protocol\cite{Liao2014,Saglamyurek2018,Rastogi2019}. The light-matter interaction under such a condition is non-adiabatic which involves the transfer between the optical coherence and the collective ground-state (spin) coherence. In the opposite case ($B < \Delta\omega_{EIT}$), the memory operation is the EIT protocol which relies on adiabatic elimination of the absorption and coherent transfer between the optical field and the collective atomic 
spin wave. The features and differences of the EIT and ATS protocols have been well studied\cite{Gorshkov2007,Saglamyurek2018, Rastogi2019}. A bandwidth of 14.7 MHz with an efficiency of 8.4 \% using the ATS protocol has been demonstrated\cite{Saglamyurek2018}. An efficiency of up to 92\% using the EIT protocol has been demonstrated but the bandwidth is only up to 2.2 MHz\cite{PhysRevLett.110.083601,PhysRevLett.120.183602,Wang2019}.  

In this paper, we explore the issues and requirements in order to extend the bandwidth of adiabatic EIT memories. Except efficiency, we define another important figure of merit for a memory, waveform likeness (WL)\cite{Zhou:12,PhysRevLett.110.083601,Wang2019}, which quantifies the degree of distortion of a probe pulse. In the single-photon regime, waveform likeness is related to the overlap of the temporal mode of photons with and without storage, which is crucial in some applications of quantum memory. We show that waveform likeness can be considered as an experimental parameter to evaluate the degree of adiabaticity of a memory. For a steady-state EIT spectrum, we identify that it can be categorized into two regimes in which the EIT transparent linewidth is proportional to the intensity or field amplitude of the control field at low and strong control field, respectively. We term these two regimes as the intensity-linear or field-linear EIT bandwidth regime. For a pulsed probe case, we show that EIT memory can be categorized into the narrowband or broadband regime, with $B < \Gamma$ or $B > \Gamma$, respectively, where $\Gamma$ is the spontaneous decay rate of the probe transition. EIT memory is limited by efficiency in the narrowband regime and is limited by waveform likeness in the broadband regime. We clarify that EIT storage can maintain a high efficiency and a high waveform likeness even in the broadband regime. The quantitative requirements to achieve such a high-performance storage are given. Our theoretical study provides essential physical insights in implementing a broadband EIT memory.   

We then present our experimental efforts toward a broadband EIT memory. We achieved a storage efficiency of $\sim 80 \%$ for probe pulses with a full-width-half-maximum (FWHM) temporal duration ($T_p$) of 30 ns, which corresponds to a bandwidth of 14.7 MHz. Limited by the available control intensity, we cannot achieve the optimized efficiency for $T_p <$ 30 ns but still obtain an efficiency of above $50 \%$ for $T_p$= 14 ns (31.4 MHz). The time-bandwidth product (TBP), defined as the ratio of the storage time at $50\%$ storage efficiency to the FWHM duration ($T_p$), is an important figure of merit for memory application. We achieve a TBP of 1267. 

\section{Theoretical study on broadband EIT storage}
The transition scheme of the $\Lambda$-type system for EIT storage is shown in Fig.~\ref{Fig1} (a). The population is assumed to be prepared in the ground state, $\ket{g}$. The weak probe field, to be stored and retrieved on demand, drives the $\ket{g} \leftrightarrow \ket{e}$ transition with a Rabi frequency of $\Omega_p$. The control field, with a Rabi frequency of $\Omega_c$, drives the $\ket{s} \leftrightarrow \ket{e}$ transition. 

Under the rotating-wave approximation, the system Hamiltonian can be described by,
\begin{equation}
\hat{H} = -\delta_{p}  \hat{\sigma_{ee}}  -\delta_{\delta_2}  \hat{\sigma_{ss}} + \frac{1}{2}(\Omega_p \hat{\sigma_{eg}} + \Omega_c \hat{\sigma_{es}} 
 + h.c.),
\end{equation}
where $\hat{\sigma_{ij}} \equiv \ket{i} \bra{j}$ denotes the atomic operator, $\delta_{p,c}$ denotes the one-photon detunings of the probe and control field, respectively, and $\delta_{2}=\delta_{p}-\delta_{c}$ denotes the two-photon detuning.

\FigOne

Theoretical analysis is based on the Maxwell-Schr\"odinger equation (MSE) and the optical Bloch equations (OBE). Under the weak-probe perturbation, the relevant equations are 
\begin{equation}
    \begin{aligned}
    &\partial_{t}{\sigma}_{eg}=(i \delta_{p}-\gamma_{ge})\sigma_{ge}+\frac{i}{2}\Omega_{c}\sigma_{sg}+\frac{i}{2}\Omega_{p} \\ 
    &\partial_{t}\sigma_{sg}=(i\delta_2-\gamma_{sg}) \sigma_{sg}+\frac{i}{2}\Omega_{c}\sigma_{eg}   
    \end{aligned}
    \label{OBE}
\end{equation}
    
\begin{equation}
    \begin{aligned}
    (\frac{1}{c}\partial_t + \partial_z) \Omega_p = i \frac{D \Gamma}{2L} \sigma_{eg},
    \end{aligned}
    \label{Maxwell}
\end{equation}
where $D$ and $L$ denotes the optical depth and the length of the atomic media, respectively\cite{PhysRevLett.120.183602}. 

\subsection{Spectral Response}
Eqs. \ref{OBE} and \ref{Maxwell} can be easily solved in the frequency domain ($\omega$-space) by Fourier transformation\cite{PhysRevLett.120.183602}. The $\omega$-space probe field can be obtained as,

\begin{equation}
W_{p}(\omega,z)=W_{p}(\omega,0)\exp[ik(\omega)z],
	\label{eq2}
\end{equation}
where
\begin{equation}
\begin{aligned}
k(\omega)=\frac{\omega}{c}+\frac{D \Gamma}{4 L} \frac{\omega}{i\omega(i(\omega+\delta_p)-\frac{\Gamma}{2})+\frac{\Omega_{c}^{2}}{4}}\equiv k_0(\omega)+k_1(\omega),
\end{aligned}
	\label{eq3}
\end{equation}
, where $k_0(\omega)=\omega/c$ is the free-space wavenumber and $k_1(\omega)$ is the spectral response function of the EIT medium, $W_p(\omega,0)$ is the spectral amplitude of the input probe pulse. 
For simplicity, we assume $\delta_c= 0$ and the ideal case with $\gamma_{sg}= 0$. For the steady-state probe transmission, we can set $\omega= 0$ and obtain $T(\delta_p)= \text{Exp}(-2 Re(k(\delta_p))L)$. The FWHM bandwidth of the EIT transparent window ($\Delta \omega_{\text{EIT}}$) is one important parameter and can be obtained as,
\begin{equation}
\Delta\omega_{\text{EIT}}=\sqrt[]{\frac{D-\ln2+2\ln2(\Omega_{c}^2/\Gamma^2)-\sqrt[]{g(D,\Omega_{c})}}{2\ln2}}\Gamma
	\label{dw},
\end{equation}
where
\begin{equation}
g(D,\Omega_{c})=(D-\ln2)(D-\ln2+4\ln2(\Omega_{c}^2/\Gamma^2))
	\label{g}.
\end{equation}
The relationship between $\Delta \omega_{\text{EIT}}$ and $\Omega_c$ is depicted in Fig.\ref{Fig2} (a). In the weak-control regime, the EIT transparent bandwidth is\cite{PhysRevLett.120.183602}, 
\begin{equation}
\Delta\omega_{\text{EIT}} \approx \sqrt{\text{ln2}}\frac{\Omega_c^2}{\sqrt{D}\Gamma}.
\label{EITbandwidth}
\end{equation}
The EIT transparent bandwidth is linearly proportional to $\Omega_c^2$ or the control intensity such that we term it as the intensity-linear bandwidth regime. In the strong-control regime, $\Delta \omega_{\text{EIT}} \approx \Omega_c$, which scales linearly with the control field amplitude such that we term it as the field-linear bandwidth regime. By setting $\Delta\omega_{\text{EIT}}$ equal to each other for both regimes, the transition $\Omega_c$ between these two regimes can be estimated to be $\Omega_c \approx 1.7\sqrt{D}\Gamma$. This explains the trend in Fig.\ref{Fig2} (a) why the intensity-linear regime is wider for a larger optical depth. It should be pointed out that intensity-linear or field-linear regime does not necessarily correspond to the EIT or ATS protocol for the memory\cite{PhysRevA.81.053836}. As mentioned in the introduction, the relative ratio of the bandwidth of the probe pulse ($B$) and the EIT transparent bandwidth ($\Delta\omega_{\text{EIT}}$) determines the EIT or ATS regime. Also, the intensity-linear or field-linear bandwidth regime does not necessarily determines the broadband or narrowband EIT storage regime either. The broadband or narrowband EIT storage depends on the relative ratio of $B$ and $\Gamma$, as will be explained later. 

\subsection{Pulse\label{sec:pulse}}
In real applications, one needs to store optical probe pulses of a finite bandwidth, instead of of a continuous wave. We consider the probe pulses with a temporal Gaussian profile with an intensity FWHM duration of $T_p$. In the frequency domain, the Rabi frequency of the probe pulse reads, 
\begin{equation}
    W_{p}(\omega,0)=\frac{T_{p}\Omega_{p0}}{\sqrt[]{4\ln2}}\exp(-\frac{\omega^2 T_{p}^2}{8\ln2}),
    \label{Wp}
\end{equation}
where $\Omega_{p0}$ is the peak Rabi frequency of the input probe pulse in time domain. The FWHM bandwidth $B$ of the probe pulse in $\omega$-space is $B = \frac{4ln2}{T_p}$. We consider the on-resonance case for the probe and control fields ($\delta_p = \delta_c = 0$) and the ideal case with $\gamma_{sg} = 0$ and $\gamma_{ge}= \frac{\Gamma}{2}$. The general case can be referred to Ref. \cite{PhysRevLett.120.183602}. 

Using Eqs \ref{eq2}, \ref{eq3}, and \ref{Wp} and Fourier transform the probe pulse back to the time domain, the solution of the slow light pulse in the time domain can be obtained as \cite{PhysRevLett.120.183602}, 
\begin{equation}
    \Omega_p(t,z=L)=\frac{1}{\sqrt{2\pi}}\frac{\Omega_{p0}T_p}{\sqrt{4\text{ln2}}}\int_{-\infty}^{\infty}d\omega e^{-i\omega t}W_p(\omega,0)e^{ik(\omega)L}.
    \label{slowlight}
\end{equation}
Applying Taylor expansion of $k_1(\omega)$ with respect to $\omega$, one obtains the dispersion terms of an EIT medium,
\begin{equation}
    \begin{aligned}
    k_1(\omega) = \sum_n a_n \omega^n\\
    a_1=\frac{iD\Gamma}{\Omega_c}, a_2=-\frac{2D\Gamma^2}{\Omega_c^4},\\
    a_3=\frac{4iD\Gamma(\Omega_c^2-\Gamma^2)}{\Omega_c^6},
    a_4=-\frac{8D\Gamma^2(2\Omega_c^2-\Gamma^2)}{\Omega_c^8},
    \end{aligned}
    \label{taylor-wl}
\end{equation}
where the explicit form of dispersion terms up to $O(\omega^4)$ are written down. The dispersion term $O(\omega)$, $O(\omega^2)$ and $O(\omega^3)$ is related to the group delay, pulse broadening, and pulse asymmetry, respectively\cite{Yu2006,PhysRevLett.120.183602}. The ratio of the absolute value of O($\omega^3$) to O($\omega^2$) for a pulse of bandwidth $B$ is, 
\begin{equation}
    |\frac{O(\omega^3)}{O(\omega^2)}|=\frac{|\Omega_c^2-\Gamma^2|}{\Omega_c^2}\frac{B}{\Gamma}.
\end{equation}
If this ratio is negligible then one can keep up to $O(\omega^2)$ term only and obtain an analytic form of the probe pulse solution, which reads,
\begin{equation}
    \Omega_p(t,z=L)=\frac{\Omega_{p0}}{\beta} Exp[-2\textrm{ln2}(\frac{(t-T_d)}{\beta T_p})^2],
\end{equation}
where
\begin{equation}
    \beta=\sqrt{1+\frac{16\textrm{ln2}D\Gamma^2}{T_p^2\Omega_c^4}},
\label{beta}    
\end{equation}
and
\begin{equation}
    T_d=\frac{L}{v_g}=\frac{L}{c}+\frac{D\Gamma}{\Omega_c^2},
    \label{Td}
\end{equation}
where $v_g$ is the group velocity of the probe pulse in the EIT medium. It should be noted that the term $L/c$ in $T_d$ may not be negligible in the short pulse regime. The energy transmission of the probe pulse is,
\begin{equation}
    T=\frac{1}{\beta}=\frac{1}{\sqrt{1+16\textrm{ln2}\frac{\eta^2}{D}}},
    \label{Trans}
\end{equation}
where 
\begin{equation}
    \eta\equiv\frac{D\Gamma}{\Omega_c^2T_p}.
    \label{eta}
\end{equation}
The parameter $\eta$ is approximately equal to $T_d/T_p$ if $v_g \ll c$. In general case, the slow light transmission efficiency (TE) can be exactly calculated by,
\begin{equation}
    \text{TE} = \frac{\int_\omega d\omega |W_p(\omega,0)|^2 exp(-2 z\text{Im}[k(\omega)]) }{\int_\omega d\omega |W_p(\omega,0)|^2 }.
    \label{TE}
\end{equation}

\FigTwo

We focus our discussion on EIT memory in the high optical depth regime (e.g. $D > 100$), where $\Omega_c \gg \Gamma$ is valid and thus $|O(\omega^3)/O(\omega^2)| \approx \frac{B}{\Gamma}$. In the narrowband regime (i.e. $B \ll \Gamma$), the formulation for keeping the dispersion up to the $O(\omega^2)$ term is already a good approximation. On the other hand, in the broadband EIT storage regime ($B > \Gamma$), $O(\omega^3)$ term could be larger than $O(\omega^2)$ term and should be considered, which may lead to the pulse distortion. We next consider the contribution of even higher order dispersion terms in the broadband EIT regime. We note that the ratio of the higher order even and odd dispersion terms, $O(\omega^{2n+2})/O(\omega^{2n})$ and $O(\omega^{2n+3})/O(\omega^{2n+1})$ with $n=1, 2, 3 ...$, are $\approx (\frac{B}{\Omega_c})^2$. Since we consider the storage in the EIT protocol, the condition $B< \Delta\omega_{EIT}$ is valid. From Fig. \ref{Fig2} (a), we know that $\Delta\omega_{EIT} < \Omega_c$ is always valid. Thus, $(\frac{B}{\Omega_c})^2 < (\frac{B}{\Delta\omega_{EIT}})^2 < 1$ is valid. Neglecting the dispersion term of $O(\omega^4)$ and above is a reasonable approximation.  

In the EIT memory protocol, one adiabatically turns off the control field to convert the optical probe field into collective atomic spin wave and then retrieve the optical field back by turning on the control field on demand. In the dark-state polariton picture, the storage and retrieval process can be considered as a coherent conversion between the optical field and atomic spin wave\cite{PhysRevLett.84.5094}. However, even if the control field is non-adiabatically (instantly) turn off and on, the additional energy loss is on the order of $v_g/c$, which is very small for most of the situations\cite{PhysRevA.64.043809, PhysRevA.100.063843}. In the storing process, one needs to select a suitable $\Omega_c$ such that the group velocity is slow enough and almost the whole probe pulse can be compressed into the media\cite{PhysRevLett.120.183602}. Also, the timing when the control field is turned off needs to be suitable such that the leakage of the front and rear tails of the probe pulse are minimized \cite{PhysRevLett.120.183602}.

To quantify this leakage loss, we define an operational efficiency, termed as $F$. Then, the overall storage efficiency (SE) of a memory can be estimated through the product of the transmission efficiency (TE) of the slow light and $F$, $\text{SE} = \text{TE} \times F$. We emphasize that there may exist additional losses during storing and retrieving periods due to the finite $\gamma_{sg}$, which we set to zero for simplicity in the current discussion\cite{PhysRevA.100.063843}. In order to compress almost the whole probe pulse into the atomic medium, choosing a suitable $\Omega_c$ to be smaller than a certain value (and thus $\eta$ larger than a certain value denoted as $\eta_{store}$ through Eq. \ref{eta}) to enable $F$ approaching to near unity is the prerequisite. The value of $\eta_{store}$ should be around 2-3\cite{PhysRevLett.120.183602}. Fig.\ref{Fig2} (b) depicts the relationship of the $\text{TE}$ and $\eta_{store}$ for some given $F$s ($F=95.5\% \sim 99.5 \%$). For a smaller optical depth and thus a smaller $\text{TE}$, the broadening effect is more severe \cite{PhysRevLett.120.183602} such that $\eta_{store}$ needs to be larger to minimize the tail leakage. More quantitative details about how to calculate these curves can be referred to Appendix \ref{app: eta}. We only concern with the high SE case, e.g. $\text{SE} > \sim 80 \%$, and in that case $\eta_{store}$ ranges $\approx 1.8 \sim 2.7$. 

In Fig. \ref{Fig2} (c), we calculate the slow light transmission via Eq. \ref{TE} versus the bandwidth of the probe pulse under the constraint of $\eta=2.43$, which corresponds to $F=98.5 \%$. The storage efficiency is thus mainly determined by TE with an uncertainty of less than $1.5\%$. From Fig. \ref{Fig2} (c), TE is nearly a constant depending on the optical depth ($D$) at low pulse bandwidth. With a larger $D$, the bandwidth of constant TE extends to a wider value, roughly proportional to $\sqrt{D}\Gamma$. This trend 
of nearly a constant TE below a certain bandwidth when EIT is in the regime of intensity-linear bandwidth regime is understandable and explained below. In the intensity-linear regime, $\eta$ can be rewritten as $\eta \propto \sqrt{D}\frac{B}{\Delta\omega_{EIT}}$ by Eq. \ref{EITbandwidth} and Eq. \ref{eta}. For a fixed $\eta$ and a fixed $D$, the ratio of pulse bandwidth to EIT bandwidth is fixed such that the slow light transmission stays a constant. From Fig. \ref{Fig2} (c), one also observes that the bandwidth range of the TE plateau is larger for a larger $D$. With a constant $\eta$, a larger $D$ means that the ratio $\frac{B}{\Delta\omega_{EIT}}$ is smaller. Therefore, more frequency components of the probe pulse lie in the central region of the EIT transparent window, which leads to a larger transmission.   

\begin{figure}[t]
{\centering\includegraphics[width=8.5cm,viewport=100 0 890 430,clip]{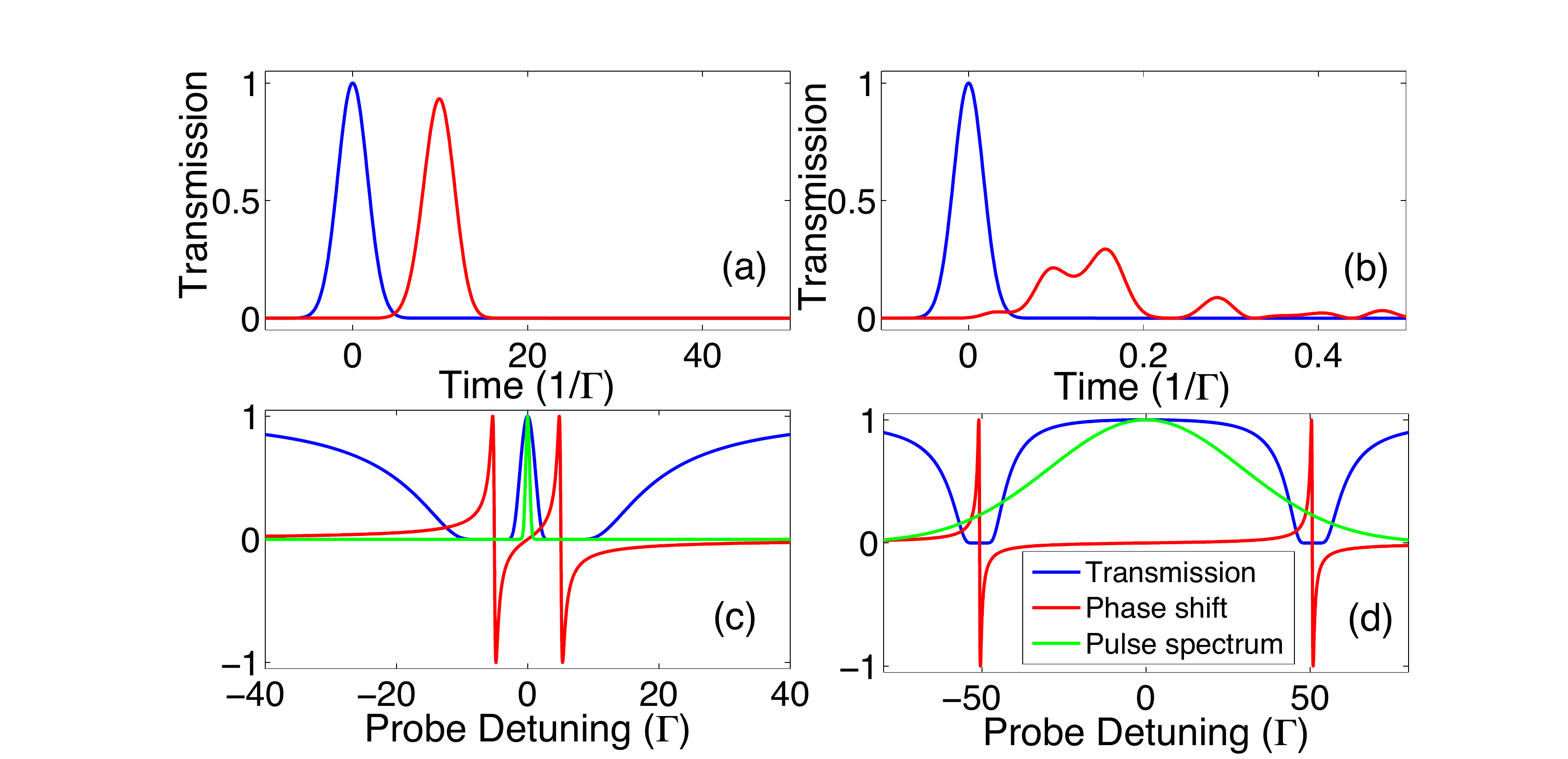}\\}
\caption{(a) and (b) the input (blue) and output probe (red) pulses with a $T_{p}$ of 4 and 0.04 $1/\Gamma$ , respectively, for $D=1000$ and $\eta=2.43$. Transmission and waveform likeness for the slow light pulse is $96.82 \%$ and $99.84 \%$ in (a) and $84.85 \%$ and $39.30 \%$ in (b), respectively. In (c) and (d), the EIT transmission spectrum, phase shift and probe pulse spectrum are plotted, corresponding to the parameters of those in (a) and (b), respectively. The maximum phase shift
	is normalized to unity for clarity.
}
\label{Fig3}
\end{figure}

In Fig.\ref{Fig2} (c), when the bandwidth of the probe pulse $B$ keeps increasing, the transmission drops and a dip structure appears and then it increases toward unity. This dip structure appears when the pulse bandwidth is on the order of the spectral separation of the two Aulter-Towens absorption peaks, which is $\sim \Omega_c$, such that its absorption is maximum. At an even higher $B$, a significant portion of the pulse spectral component are beyond the Aulter-Towens absorption peaks and lie in the far-detuned and near transparent regime. This is why the transmission rises again. Although TE is high under this circumstance, the probe pulse experiences a severe distortion because high order dispersion terms are not negligible. As an example, Fig. \ref{Fig3} (a) and (b) depicts two slow light pulses with a $T_p$ of 4 and 0.04 $1/\Gamma$, respectively, after passing through an EIT medium with $D=1000$ and $\eta=2.43$. Fig. \ref{Fig3} (c) and (d) depict the corresponding EIT transmission, phase shift and the pulse spectrum. For the long pulse case, its spectrum is relatively narrow and lies within the transparent window. Also, the pulse spectrum lies in a spectral range of nearly linear phase shift dependence on the detuning. Thus, the group velocity is well-defined and the slow light pulse resembles the input pulse with a certain group delay. For the short pulse case, however, its spectrum extends over the two Aulter-Townes absorption doublets and also lies in the complicated profile of the phase shift spectrum. The pulse not only experiences a significant absorption but also a significant dispersion such that the output pulse has a significant distortion. In the quantum memory application, this serious distortion in the temporal mode may introduce some complications in the application. Except the efficiency, it seems necessary to define another figure of merit to quantify this distortion for evaluating the performance of an optical memory.

\subsection{Waveform Likeness\label{subsec: waveform likeness}}

\begin{figure*}[t]
{\centering\includegraphics[width=17cm]{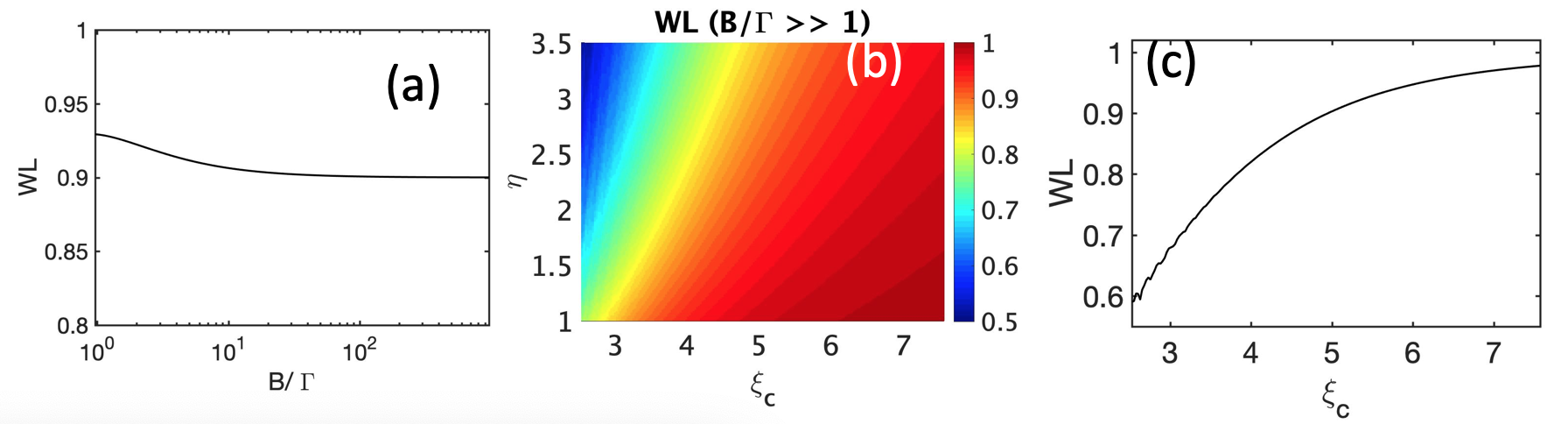}\\}
\caption{ (a) Solid curves denote simulated WLs with different bandwidths when fixing $(\xi_c, \eta) = (5.18, 2.77)$. (b) WL of different $(\xi_c, \eta)$ when $B/\Gamma \gg 1$. In the simulation, $B/\Gamma = 880$. Note that WL will converge to the value in this figure in the broadband limit. (c) The cross section of (b) along the $\xi_c$ axis for $\eta = 2.5$.}
	\label{Fig4}
	\end{figure*}

Inspired by the aforementioned discussion, we define a figure of merit, waveform likeness (WL), to quantify the extent of distortion, which reads as,
\begin{equation}
\text{WL} \equiv \frac{\left|\int_{-\infty}^{\infty} \Omega_{p,in}^{*}(t-T_d)\Omega_{p,out}(t)dt\right|^{2}}{\int_{-\infty}^{\infty}\left|\Omega_{p,in}(t)\right|^{2}dt \int_{-\infty}^{\infty}\left|\Omega_{p,out}(t)\right|^{2}dt}.
	\label{def-wl}
\end{equation}
Waveform likeness reflects the similarity between the input and slowed (or retrieved) probe pulses\cite{PhysRevLett.120.183602, PhysRevLett.110.083601}. Fig. \ref{Fig2} (d) depicts the WL of different input pulse bandwidths, corresponding to the TE of Fig. \ref{Fig2} (c). The severe pulse distortion at high bandwidth can be manifested by the degradation of WL. For clarity, we do not show the WL below $50\%$ in Fig. \ref{Fig2} (d). Similar to the TE, there is a plateau for WL at the low $B$ regime. For a larger $D$, this plateau extends to a wider range, but is slightly smaller than that of the TE. This suggests that WL is a more stringent parameter than the TE at high pulse bandwidth.

In the example shown in Fig.\ref{Fig3} (b) for the broadband case, the output pulse profile splits over a long period of time. Although TE remains high ($84.85 \%$), the WL is very low ($39.3 \%$) and this highlights the need to introduce the parameter WL. In the case of Fig.\ref{Fig3} (b), one needs to collect the trifling signals over a long time window to maintain the desired SE. It may affect the feasibility of storage, especially in the single-photon regime for quantum memory application since the noises makes it arduous to collect signal over the long time window. Besides, the group delay is not well-defined in the broadband case. A certain portion of the pulse is delayed with sufficient time but there exists a front tail virtually without any delay. It is therefore not feasible to store the probe pulse with a negligible amount of leakage, or $F \approx 1$. 

In the single-photon regime, waveform likeness is related to the overlap of the temporal mode of the output to the input pulse. A high WL is important since it is essential to preserve a good single-photon waveform to reach a high non-classical property in quantum storage\cite{Zhou:12}. In the application of quantum memory, it may involves with the Hong–Ou–Mandel (HOM) two-photon interference. The contrast of HOM interference depends on the degree of indistinguishability of the temporal \cite{Fearn:89} and spectral mode \cite{PhysRevLett.100.133601, PhysRevA.77.022312} for the two photons. The severe distortion increases the difficulty of creating two indistinguishable photons in both temporal and spectral mode. Besides, with a better WL, the arrival time of the retrieved signal can be well-controlled, instead of spreading out over a long time period. This temporal control is also important to quantum communication through time-bin protocol \cite{PhysRevA.66.062308}. In the following subsections, we present further analysis of the WL.

\subsubsection{Broadband Limit}
Examine the numerator in the definition of WL in Eq. \ref{def-wl}, one can simplify it to,
\begin{equation}
    \text{WL} \propto |\int d\omega |W_p(\omega,0)|^2e^{i(k_1(\omega)L-\frac{D\Gamma\omega}{\Omega_c^2})}|^2. 
    \label{WLint}
\end{equation}
The expression of $k_1(\omega)$ of Eq.\ref{eq3} can be written as, 
\begin{equation}
    k_1(\omega)L=\frac{D\Gamma}{\Omega_c^2}\frac{\omega}{1-2i\frac{\omega\Gamma}{\Omega_c^2}-\frac{4\omega^2}{\Omega_c^2}}.
    \label{k1}
\end{equation}
For a probe pulse of bandwidth $B$, $\frac{B}{\Omega_c} <1$ in the broadband EIT memory regime since $\Delta_{EIT} \simeq \Omega_c$. Also, we have $B > \Gamma$ such that $\frac{\Gamma}{\Omega_c} < \frac{B}{\Omega_c} < 1$. Under such conditions, the term $2\frac{\omega\Gamma}{\Omega_c^2} \ll1$ in the denominator of Eq.\ref{k1} and thus it can be neglected. In that situation, $k_1(\omega)$ is almost a real number and reads as,
\begin{equation}
    k_1(\omega)L\simeq \frac{D\Gamma\omega}{\Omega_c^2-4\omega^2}.
\end{equation}
Thus, the transmission efficiency of the output probe pulse is near unity, as can be seen from Eq. \ref{TE} because $\text{Im}(k_1(\omega))$ is near zero. This is understandable since the EIT transparent window is very wide and has a flat-top profile in the broadband regime, such as that shown in Fig. \ref{Fig3} (d). If $B \ll \Omega_c$, almost all the pulse spectrum lies in the near-unity transparent window and thus the transmission is near unity. 

Because WL is actually determined by the relative interplay between $W_p$ and $k_1(\omega)$, we can consider all spectral behaviors normalized to $B$. We define $\widetilde{k_1}(\widetilde{\omega})=k_1(\widetilde{\omega}B)$ and $\widetilde{W_p}(\widetilde{\omega})=W_p(\widetilde{\omega}B)$, where $\widetilde{\omega}=\frac{\omega}{B}$. When considering the different bandwidth $B$, $\widetilde{W_p}(\widetilde{\omega})$ remains the same and the WL of spectral integral Eq.\ref{WLint} is only determined by $\widetilde{k_1}(\widetilde{\omega})$, which reads,  
\begin{equation}
  \widetilde{k_1}(\widetilde{\omega})L=\frac{D\Gamma}{B}\frac{\widetilde{\omega}}{(\frac{\Omega_c}{B})^2-4\widetilde{\omega}^2}.
  \label{k1norm}
\end{equation}
Here, we introduce two parameters: $\xi_c$ and $\xi_D$, defined as
\begin{equation}
    \begin{aligned}
    \xi_c \equiv \frac{\Omega_c}{B}\\
    \xi_D \equiv \frac{D\Gamma}{B}.
    \end{aligned}
    \label{xicd}
\end{equation}
Therefore, Eq.\ref{k1norm} can be rewritten
\begin{equation}
\begin{aligned}
\widetilde{k_1}(\widetilde{\omega}) L \approx \xi_D   \frac{\widetilde{\omega}}{\xi_c^2-4\widetilde{\omega}^2}
\end{aligned}
	\label{k-eta-xi}
\end{equation}

If we keep $\xi_c$ and $\xi_D$ stay at the same values when varying $B$, the normalized response, $\widetilde{k_1}(\widetilde{\omega}) L$, can be maintained exactly at the same value through adjusting the $\Omega_c$ and $D$ according to Eq. \ref{xicd} with $\Omega_c = \xi_c B$ and $D = \xi_D \frac{B}{\Gamma}$. Since the same normalized spectrum maintains the same WL, the values of $\xi_c$ and $\xi_D$ therefore determine the criterion for the waveform likeness. These two parameters connect to $\eta$ by $\eta = \frac{1}{4 ln2} \frac{\xi_D}{\xi_c^2}$. For simplicity, we consider hereafter $\xi_c$ and $\eta$ as the two independent variables, while $\xi_D$ can be determined by the other two parameters.
Thus, the whole picture of WL in the broadband limit can be summarized by $\eta$ and $\xi_c$. It should be pointed out that $\xi_c$ denotes the ratio of EIT bandwidth to the bandwidth of input probe pulse in the broadband EIT regime. Conceivably, a larger $\xi_c$ means that the probe spectrum is well-located in the central EIT window, which results in a high WL. 

Note that once fixing $(\xi_c, \eta)$, WL converges to a constant value in the large bandwidth limit, as depicted in the example of Fig.\ref{Fig4} (a). An numerical simulation of the convergent WL for different $(\xi_c, \eta)$ with a given large $B$ of 880$\Gamma$ is shown in Fig.\ref{Fig4} (b). In the broadband EIT regime, one can use Fig.\ref{Fig4} (b) to determine WL and thus estimate the required experimental parameters, ($D, \Omega_c$) by Eq. \ref{xicd}. In other words, under the condition of $B/\Gamma \gg 1$, all information about WL has been mapped in Fig.\ref{Fig4} (b). The trend of WL can be easily captured by Fig.\ref{Fig4} (b): decreasing $\eta$ and increasing $\xi_c$ is favorable to WL. However, $\eta$ has a minimum requirement such that the storage of almost the whole pulse is possible. For a fixed $\eta$ of 2.5, WL is determined solely by $\xi_c$, as depicted in Fig.\ref{Fig4} (c). In the high-WL condition, the pulse distortion majorly comes from the third order dispersion term. Under such a condition for WL $\approx$1, the approximate analytical form of WL can be obtained, which reads, 
\begin{equation}
    \text{WL} {\approx} (1 - 5.4 \times (\frac{\eta}{\xi_c^2})^2)^2.
    \label{WL-analy}
\end{equation}
With a given parameter set of $(\xi_c, \eta)$, the approximate WL is also determined.

	\begin{figure}[t]
	{\centering\includegraphics[width=9.5cm]{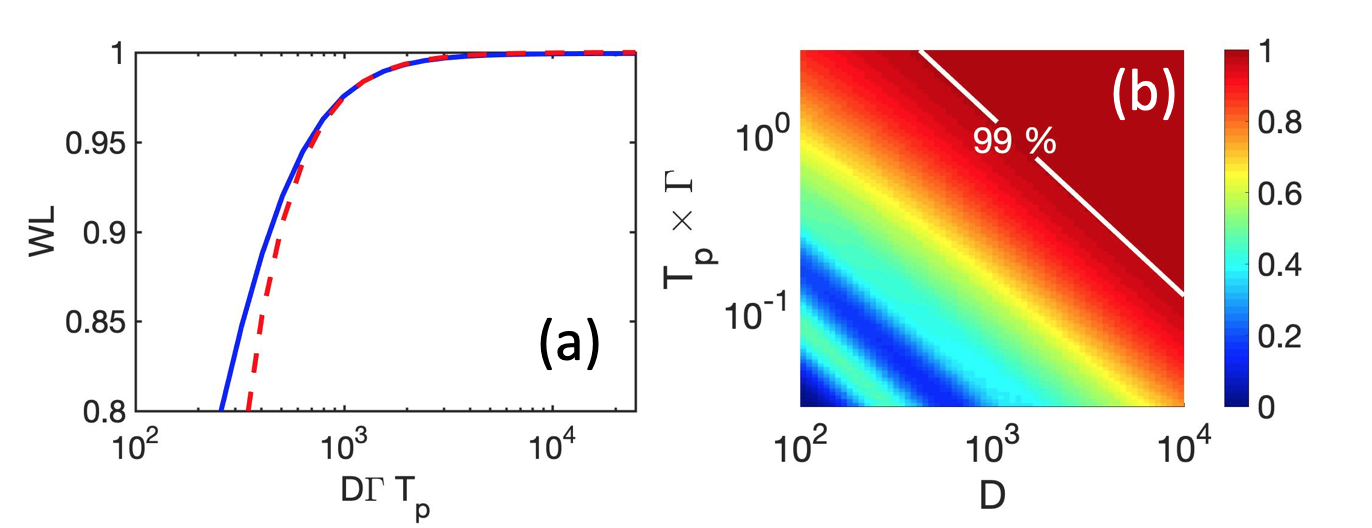}\\}
	\caption{ (a) WL versus $D T_p \Gamma$. Here, we fix $D = 1000$ and vary $T_p$. The solid curve is based on the numerical simulation and the dashed curve is calculated through Eq.\ref{WL-analy}. (b) WL as a function of $T_{p}$ and $D$ in logarithmic-logarithmic scale. In both graphs, $\eta = 2.5$ is assumed.
	}
	\label{Fig5}
	\end{figure}

\subsubsection{Adiabatic Storage}
Experimentally, the three variable parameters are $D, \Omega_c$ and $T_p$(or $B$). By putting the relation of $\xi_c$ into that of $\eta$ and reducing $\Omega_c$, one gets, 
\begin{equation}
    D T_p \Gamma = \eta \xi_c^2 \times (\text{4ln2})^2. 
    \label{DTG}
\end{equation}
In Ref.\cite{PhysRevA.76.033805}, the parameter $D T_p \Gamma$ has been identified as a measure of degree of adiabaticity in EIT storage and a larger value in this parameter indicates a higher degree of adiabaticity. Therefore, we term it as the adiabaticity parameter. As mentioned before, the parameter $\eta$ has to be larger than a certain value (e.g. $\sim$2.5) in order to store almost the whole probe pulse into the media. The adiabaticity parameter $D T_p \Gamma$ is thus mainly determined by the parameter $\xi_c$. Once both $\eta$ and $\xi_c$ are determined, the adiabaticity parameter also determine the WL through Eq. \ref{WL-analy}.   

The dash line in Fig.\ref{Fig5} (a) depicts an example of the calculated WL through Eq. \ref{WL-analy} versus the adiabaticity parameter for $D=1000$ and $\eta=2.5$. The solid line is WL from the numerical calculation. The analytic formula matches the numerical calculation at high WLs but shows certain deviation at lower WLs.  Fig.\ref{Fig5} (b) depicts a numerical calculation of the two-dimensional contour plot of WL versus $T_p\Gamma$ and $D$ for $\eta=2.5$. It is evident to observe that WL is nearly a constant for a constant product of $D T_p \Gamma$. This is understandable since the adiabaticity parameter $D T_p \Gamma$ is related to the parameter set $(\xi_c, \eta)$ by Eq. \ref{DTG}. And a given parameter set of $(\xi_c, \eta)$ directly determine the WL through Eq. \ref{WL-analy}. As mentioned, the parameter $D T_p \Gamma$ can be considered as a measure of the degree of adiabaticity\cite{PhysRevA.76.033805}. Due to its direct relation with WL, one can reverse the logic and consider WL as an experimentally observable parameter useful for evaluating the degree of adiabaticity for the memory.  

In comparison with the ATS protocol \cite{Saglamyurek2018, PhysRevA.100.012314}, which utilizes the non-adiabatic process to convert the polarization coherence into spin coherence, the value $D T_p \Gamma$ locates in $1 \sim 100$ and therefore the WL is not high. In the ATS protocol, the pulse distortion comes from the oscillation between the polarization and spin coherence and one extracts one of the pulse out but not all of the photonic signal in the temporal domain. If one only focus on the extracted pulse, it is not severely distorted but remains a Gaussian shape \cite{Saglamyurek2018, PhysRevA.100.012314}. But it certainly sacrifices the efficiency. In the EIT protocol, we include all optical signal in the temporal domain in the evaluation of WL. It is possible to reach a high WL and a high SE simultaneously for adiabatic EIT storage. However, there is a high demand on $D$ and $\Omega_c$ for EIT protocol, as will be discussed below. 

\subsection{Experimental Requirements \label{sec: multiple requirements}}

	\begin{figure}[t]
	{\centering\includegraphics[width=8.7cm,viewport=60 0 890 440,clip]{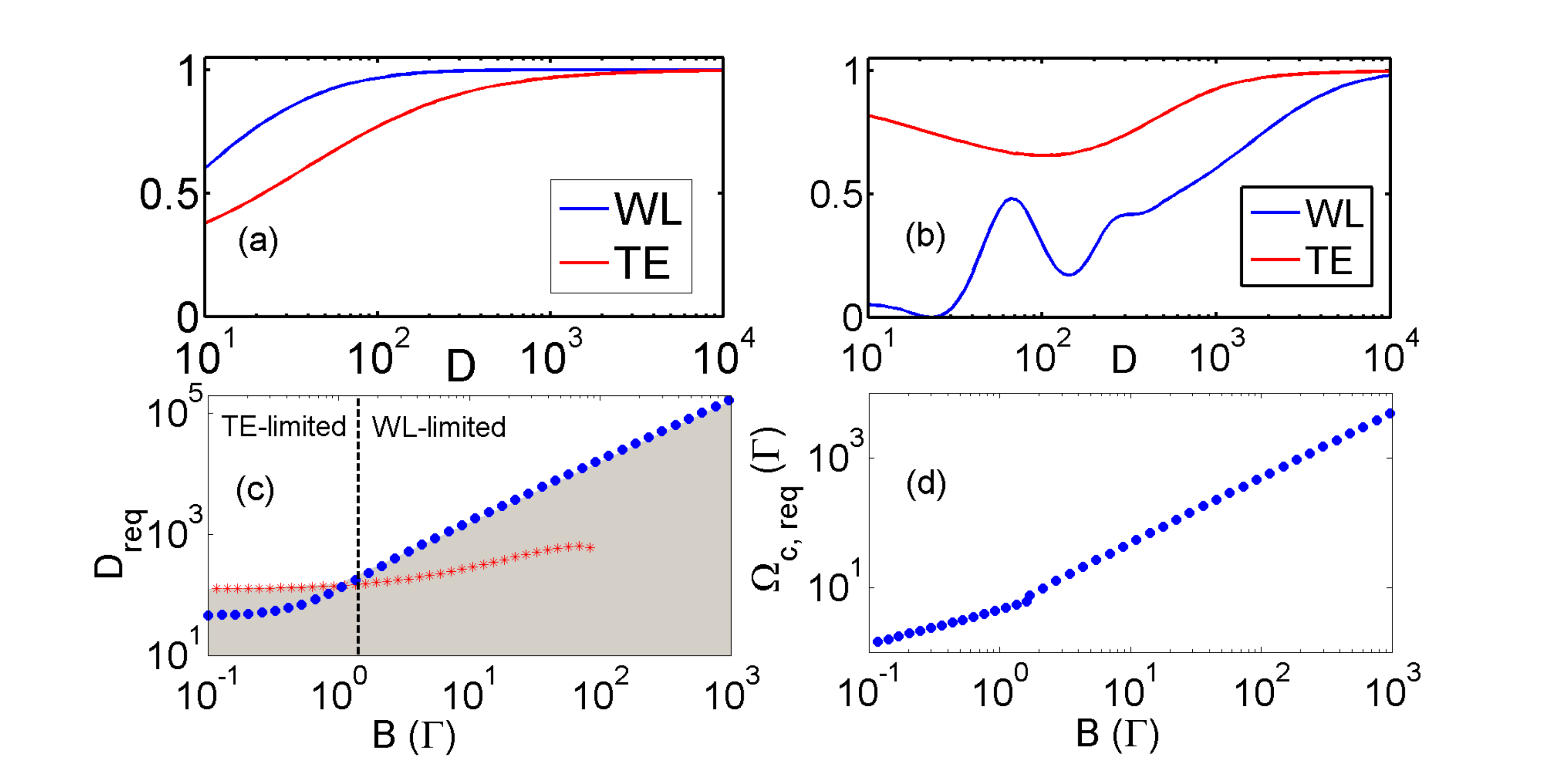}\\}
	\caption{ (a) and (b) depicts the TE and WL in a given $D$ and $T_p \Gamma = 10$ ($0.004$) or $B/\Gamma \approx 0.277$ ($693.15$). Here, $\Omega_c$ is determined by the relation $\eta = 2.5$. (c) The required optical depth ($D_{\text{req}}$) versus the bandwidth of probe pulse ($B$). The red (blue) markers denote the required optical depth to satisfy TE $\geq 80\%$ (WL $\geq 90\%$). Optical depth above the shaded area can satisfy both TE $\geq 80\%$ and WL $\geq 90\%$. (d) The corresponding required $\Omega_c$ for $D_{\text{req}}$ in c, which meets both TE $\geq 80\%$ and WL $\geq 90\%$.
	}
	\label{Fig6}
	\end{figure}

In this section, we want to provide a guide on the required optical depth $D$ and control intensity (characterized by $\Omega_c$) when implementing a broadband EIT memory with a value of SE and WL above a given threshold. Fig. \ref{Fig6} (a) and (b) depict two representative examples of TE and WL versus $D$ for the narrowband and broadband cases with $B/\Gamma$ of 0.277 and 693.15, respectively, with $\eta=2.5$ to satisfy the storage requirement. In the narrowband case, the WL is larger than the TE, while the situation is opposite for the broadband case. In other words, TE (WL) is the bottleneck when operating the narrowband (broadband) EIT storage. From these figures, one can determine the minimum required optical depth in order to obtain the TE and WL of larger than a certain value (e. g. with TE $\geq 80\%$ and WL $\geq 90\%$) simultaneously. From Eq. \ref{eta}, one can then determines the required $\Omega_c$ for a fixed $\eta$. Considering the similar calculations for different bandwidth $B$, one can reach Fig. \ref{Fig6} (c), which depicts the required $D$ versus $B$ for the request with TE$\geq 80\%$ and WL$\geq 90\%$. The optical depth above the shaded region will satisfy both the threshold conditions on TE and WL. It is evident that the required optical depth (denoted as $D_{req}$) is constrained by TE or WL when $B \lesssim \Gamma$ or $B \gtrsim \Gamma$, respectively. We term both regimes as the TE-limited or WL-limited regime, respectively. This trend is related to the fact mentioned in section \ref{sec:pulse} that one has to consider the dispersion up to O($\omega^3$) for the broadband EIT memory regime ($B > \Gamma$) but is enough to O($\omega^2$) for the narrowband ($B < \Gamma$) regime. 

The TE-limited feature in the narrowband EIT regime also explains why the previous researches for narrowband EIT memories did not bothered by the distortion issue since SE is a bottleneck in that regime, not the WL\cite{PhysRevLett.110.083601,PhysRevLett.120.183602}. On the contrary, upon pushing towards the broadband regime, one steps into the WL-limited regime and the pulse distortion is a must to consider.
From Fig. \ref{Fig6} (c), it can be observed that the required optical depth is linearly proportional to the bandwidth in the WL-limited regime. This trend is understandable if we look at Eq. \ref{DTG} and modify it to,
\begin{equation}
    D_{req}=\text{4ln2}\eta\xi_c^2\frac{B}{\Gamma}.
\end{equation}
Fig. \ref{Fig6} (d) depicts the required $\Omega_c$ to meet the request on TE and WL. Clearly, the demand on $D$ and $\Omega_c$ is very high for EIT protocol to achieve a high TE and WL. As a reference, a $B$ of $\sim 10 (100) \Gamma$ requires $D\sim 10^3 (10^4)$ and $\Omega_c \sim 10^2 (10^3)$ for a WL$\geq 90\%$. Experimentally, an optical depth of larger than $10^3$ has been achieved for cold atoms in free space or in a hollow-core fiber\cite{PhysRevA.90.055401}. With cold atoms inside a cavity, an effective optical depth of 7600 has been achieved\cite{Ducavity}. Another way to achieve a high TE and a high WL using the EIT protocol but with a much lower optical depth is to utilize the forward storage and backward retrieval scheme\cite{PhysRevLett.98.123601,PhysRevLett.110.083601}, although some experimental complexities are needed. 

While the broadband EIT memory requires a high optical depth and a strong control intensity, we expect that some complicated effects may appear in realistic situation which are out of the scope of this paper. For example, nonlinear optical effect such as the photon switching effect due to the off-resonant excitation of the coupling field to the nearby transition \cite{Schmidt:96,PhysRevLett.81.3611} or the four-wave mixing \cite{PhysRevA.88.013823} may become a serious issue. But there are also some methods that can reduce these effects\cite{PhysRevLett.120.183602}. The influence of cooperative effect due to the resonant dipole-dipole interaction on the EIT memory may become significant and await further study\cite{PhysRevLett.120.183602}. 

With these theoretical studies in mind, we then present our experimental efforts towards broadband EIT memory.

\section{Experimental Setup}

\begin{figure}[t]
\includegraphics[width=8.5cm,viewport=75 20 780 590,clip]{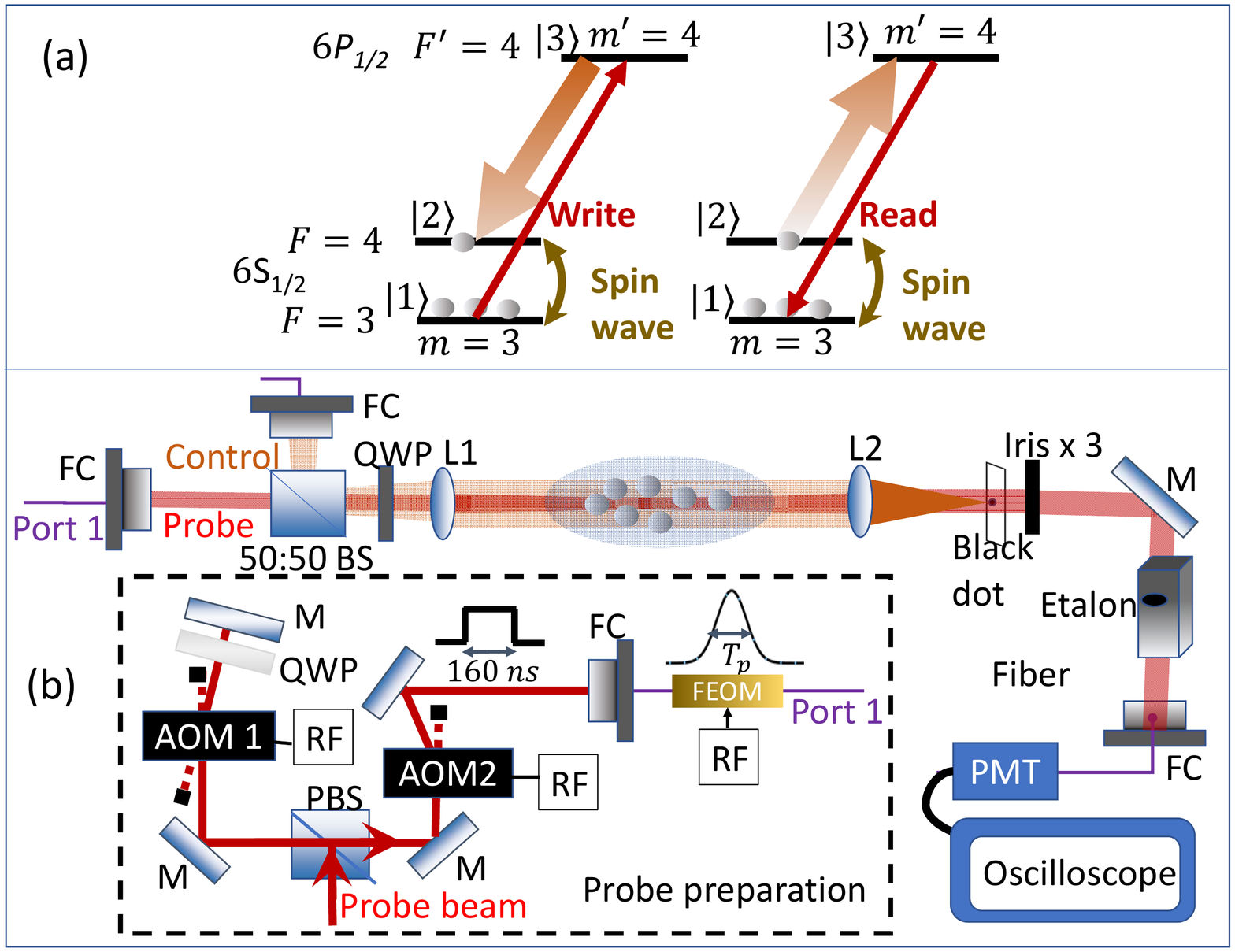}
\caption{ (a) Energy levels and laser excitations of $^{133}Cs$ for EIT memory. (b) Experimental setup for EIT memory. AOM: acousto-optic modulator; BS: beam splitter. FEOM : fiber electro-optic modulator; RF: radio-frequency signal; PMT: photomultiplier tube; M: mirror; L: lens; QWP: quarter waveplates; FC: fiber coupler; Port 1 is transmitted by a polarized-maintaining fiber.}
\label{setup}
\end{figure}

We utilize a cesium magneto-optical trap (MOT) with a cigar-shaped atomic cloud to implement the EIT-based optical memory. To increase the optical depth of atomic media, we employ the temporally dark and compressed MOT, as well as Zeeman-state optical pumping\cite{PhysRevA.90.055401}. Pumping population towards the single Zeeman substate ($| F=3, m=3 \rangle$) also makes the storage performance less sensitive to the stray magnetic field \cite{Peters:09, PhysRevLett.120.183602}, which is desirable for the long-time storage. Efforts are made to reduce the ground-state decoherence rate $\gamma_{21}$, such as minimizing the stray dc and ac magnetic fields and using the near co-propagating probe and control beams to reduce the residual Doppler broadening. Details of the MOT setup can be refereed to\cite{PhysRevA.90.055401,PhysRevLett.120.183602}. 

The EIT optical memory is operated at the $D_1$ line with the probe beam driving the $|F=3\rangle \rightarrow |F'=4\rangle$ $\sigma^+$-transition and control beam driving the $|F=4\rangle \rightarrow |F'=4\rangle$ $\sigma^+$-transition, as shown in Fig.\ref{setup} (a). As pointed out in \cite{PhysRevLett.120.183602}, operating the EIT at $D_1$ transition can reduce the control-intensity-dependent ground-state decoherence rate due to the off-resonant excitation of the control beam to the nearby transition. The detailed setup is plotted in Fig.\ref{setup} (b). The probe beam from a laser source doubly passes one acousto-optic modulator (AOM1) for adjusting its frequency with a minimal spatial movement. It is then sent to another AOM (AOM2) for switching with a 160-ns square pulse. The probe beam is then coupled into a fiber electro-optic modulator (EOM) to shape the probe pulse into a Gaussian waveform with a FWHM of larger than 10 ns. Due to the finite extinction ratio ($\sim$ 18 dB) of the fiber EOM, adding AOM2 as an additional switch is to minimize the probe leakage during storage. The probe beam is then coupled with the control beam through a $50:50$ beam splitter. Both beams are sent into the cold atomic ensembles. Before coming into MOT cell, the probe beam is focused by a lens (L1) to an intensity $e^{-2}$ diameter of $\sim$100$\mu$m around the atomic clouds while the coupling beam is collimated by the same len (L1) with a diameter of $\sim$ 240 $\mu$m. After going out from the MOT cell, the control beam is focused by another lens (L2) and is then blocked by a black dot. Probe beam, on the other hand, is collimated by the lens L2 and then coupled into a fiber before passing through three irides and an etalon filter, which filters out unwanted control light upon detection. The probe beam is then detected by a photomultiplier tube (Hamamatsu R636-10).

\section{Results and Discussions}
Here, we present our experimental results on broadband storage using EIT protocol. We first discuss the efficiency versus the optical depth at a fixed temporal width for the input probe pulse. Then we vary the temporal width (or bandwidth) of the input probe pulse and study its efficiency dependence. Finally, we study the efficiency dependence on the storage time.  

\begin{figure}
\includegraphics[width=8.6cm,viewport=85 10 900 440,clip]{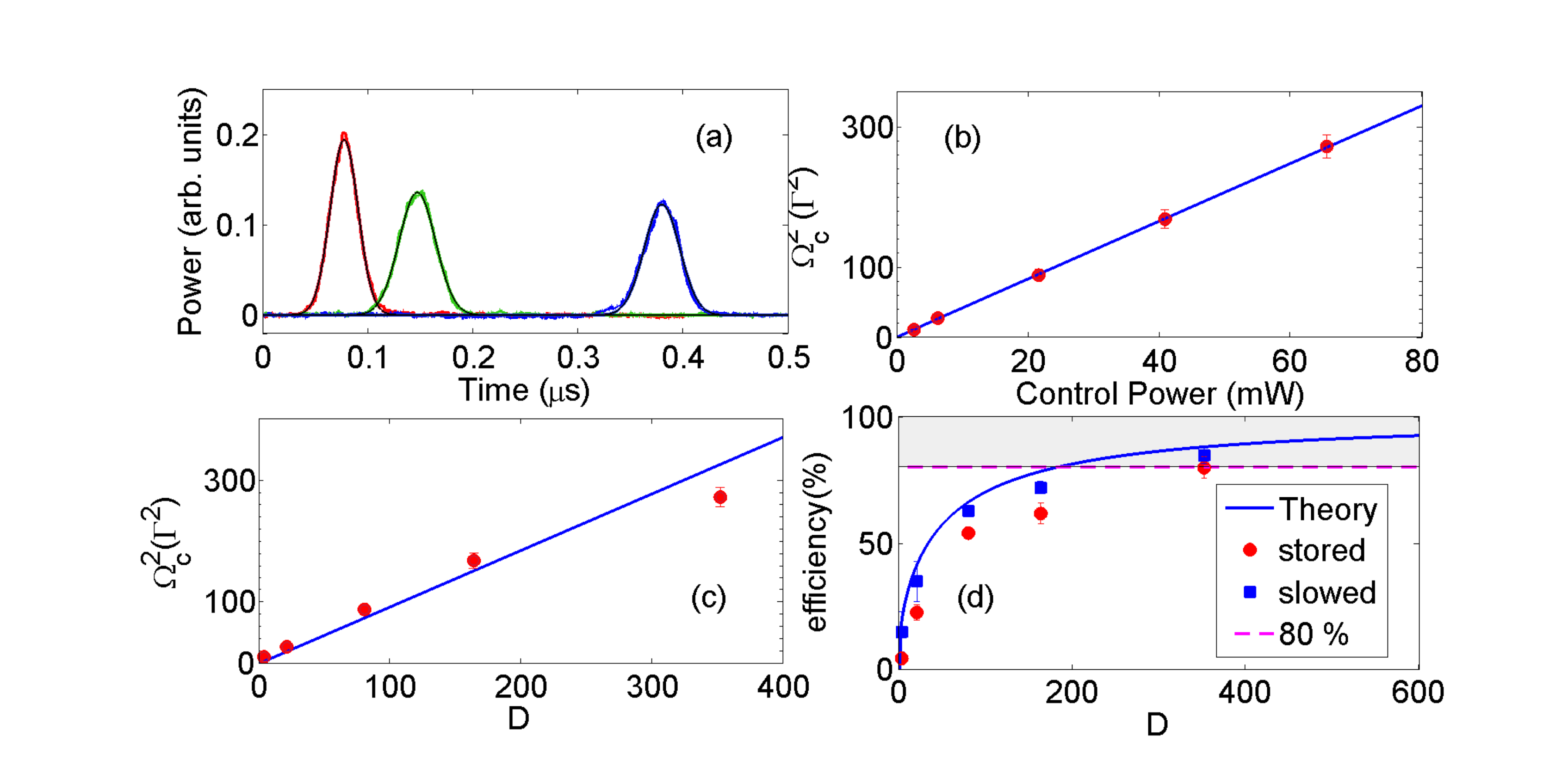}
\caption{In (a), the red, green and blue curves represent the input, slowed (efficiency $86.1 \%$), and stored-and-retrieved (efficiency $82.0 \%$) probe pulses. The parameters $D$ and $\Omega_c$ are 392 and 16.48 $\Gamma$, respectively. (b) depicts the $\Omega_c^2$ versus the control power. The solid blue line is a linear fit to the data. (c) depicts the optimum $\Omega_c^2$ which maximize the efficiency versus the optical depth. The blue solid line is a linear fit to the data. (d) depicts the efficiencies of the slowed (blue square) and stored-and-retrieved (red circle) probe pulses versus the optical depth. The solid blue line is a theoretical calculation of the slow light efficiency, assuming $\gamma_{gs} = 0$, $\gamma_{ge} = 0.75 \Gamma$ and $\eta$= 2.5.}
\label{diff_D}
\end{figure}

\subsection{Efficiency dependence on optical depth}
In a previous work\cite{PhysRevLett.120.183602}, we have studied the storage efficiency versus the optical depth for probe pulses of $T_p=$ 200 ns. Because the pulse bandwidth $B=2\pi\times$ 2.2 MHz is smaller than the spontaneous decay rate of $\Gamma=2\pi\times$ 4.56 MHz, the storage was in the narrowband regime. To entering the broadband regime, we firstly consider the storage of probe pulses with a $T_p$ of 30 ns, corresponding to a bandwidth of $B=2\pi\times$14.71 MHz. We vary the optical depth ($D$) and adjust the control intensity for each $D$ to obtain an optimized efficiency. A representative raw data showing the input, slowed, and stored-then-retrieved probe pulses is shown in Fig. \ref{diff_D} (a). In this case, the parameters $D$ and $\Omega_c$ are 392 and 16.48 $\Gamma$, respectively. The $\Omega_c$ is determined by the spectral separation of the Aulter-Townes splitting in the EIT spectrum, taken at a very low optical depth (e. g. $D < 3$) such that the Aulter-Townes doublets are clear. The optical depth is determined by the spectral fitting of the probe transmission spectrum with the control field off. In the fitting, we set $\gamma_{ge}= 0.75 \Gamma$ which takes the finite laser linewidth and the laser frequency fluctuation into account\cite{PhysRevLett.120.183602}.  

As a consistency check of the determined $\Omega_c$, the optimized $\Omega_c^2$ for each $D$ versus the control power is shown in Fig. \ref{diff_D} (b). The data fit very well to a linear relation as expected. Fig. \ref{diff_D} (c) depicts the optimized $\Omega_c^2$ versus $D$, which reasonably follows a linear relation. We roughly choose $\eta$ to be nearly a constant to obtain the optimized efficiency.  Due to Eq.\ref{eta}, this implies that $\Omega_c^2$ is linearly proportional to $D$ for a fixed $T_p$. The corresponding efficiencies of the slowed and stored-then-retrieved probe pulses are shown in the blue squares and red circles of Fig. \ref{diff_D} (d), respectively. The blue solid line is a theoretical calculation of the slow light efficiency based on Eq. \ref{Trans} with $\gamma_{ge}= 0.75 \Gamma$ and $\eta=0.25$. The theoretical curve matches well with the slow light efficiencies. At the highest $D$ of 392, the achieved efficiency of the stored-then-retrieved pulse is $79.65(1.50)\%$.   

\subsection{Efficiency dependence on the temporal width}

\begin{figure}
\includegraphics[width=8.6cm,viewport=90 5 900 440,clip]{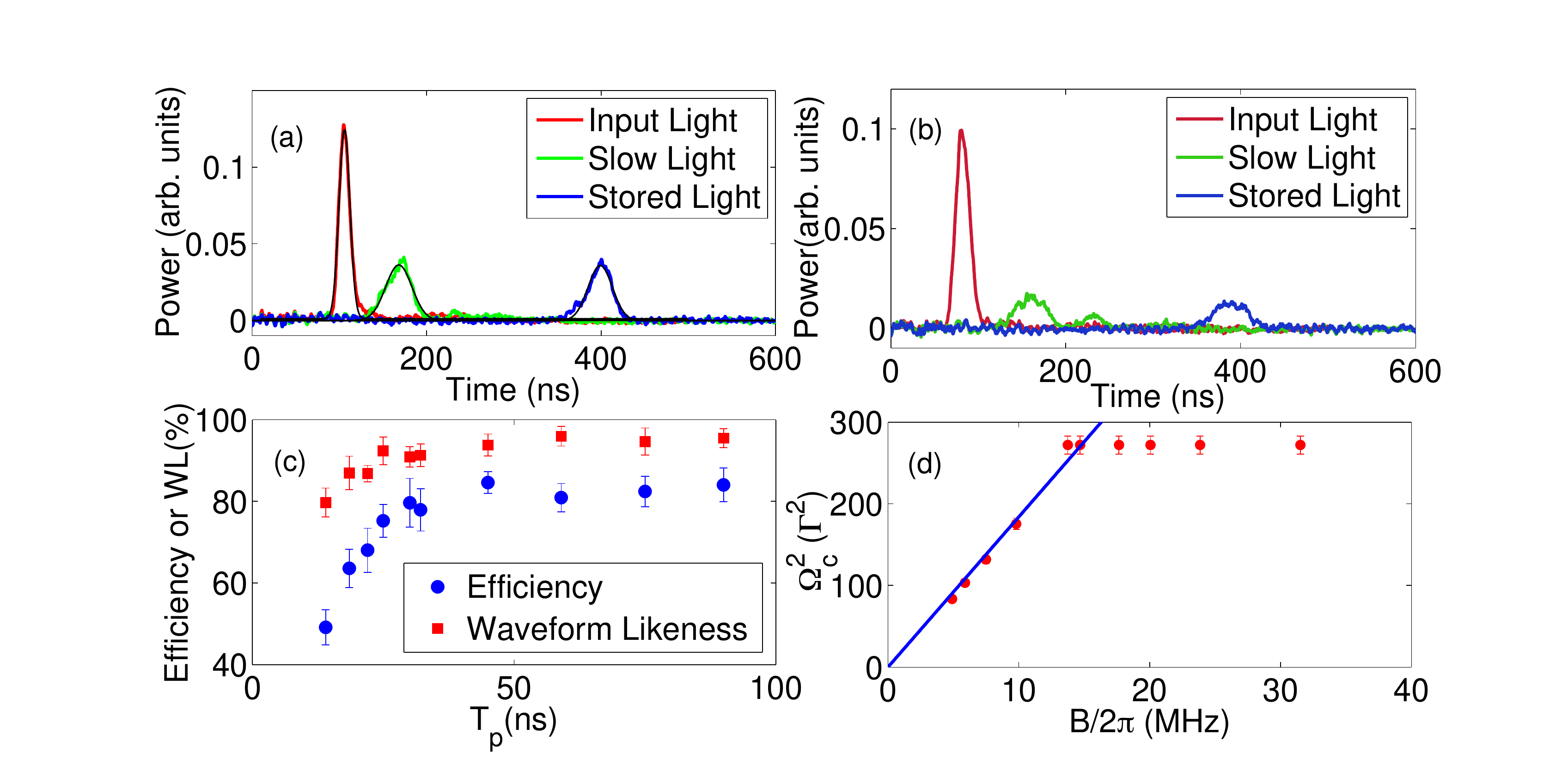}
\caption{ Fig. (a) depicts one example of the raw data with $T_p= 14 $ns, $D=356$ and $\Omega_c=17.37\Gamma$. The efficiency and waveform likeness for the stored-then-retrieved pulse is $54.8\%$ and $79.7 \%$, respectively. Fig. (b) depicts one example of the raw data to demonstrate the distortion in the slow light pulse. In this case, the parameters $\{T_p, D, \Omega_c\}$ are $\{18.6 \text{ns}, 123, 8.3 \Gamma\}$, respectively. The efficiency and WL for the stored-then-retrieved pulse are 31.8 \% and 64.7\%, respectively. (c) The efficiency (blue circle) and waveform likeness (red square) versus the FWHM pulse duration. For $T_p < 30$ ns, we are limited by available control power, as also shown in (d), such that the efficiency and WL goes down. (d) The used $\Omega_c^2$ versus the pulse bandwidth for the data corresponding to those of (c). The blue solid line is a linear fitting curve for the five date points with narrower bandwidth (or larger $T_p$).}
\label{diff_timewidth}
\end{figure}

Keeping at a large $D$ of 356, we then vary $T_p$ from large to small values (or $B$ from small to large values) and adjust the $\Omega_c$ to obtain an optimized efficiency. Unfortunately, the control intensity is technically limited to a certain level (corresponding to $\Omega_c= 17.37 \Gamma$) such that when $T_p < 30$ ns we cannot obtain the optimized efficiency. The results are shown in Fig. \ref{diff_timewidth} (c). The corresponding $\Omega_c^2$ versus $B$ are shown in Fig. \ref{diff_timewidth} (d). For the data with $T_p >$ 30 ns, the $\Omega_c^2$ versus $B$ is well fit by a linear relation, which means that $\eta$ is maintained at a constant value. The corresponding efficiencies for $T_p >$ 30 ns are pretty much a constant when $\eta$ is kept at a constant. The efficiency goes down when $T_p <$ 30 ns, due to the non-optimized $\Omega_c$. Based on the fit curve of the input and retrieved probe pulses, we can calculate the waveform likeness. The WL are shown in the red square data in Fig. \ref{diff_timewidth} (c). With our current experimental parameters, the EIT storage is in the TE-limited regime such that WL is larger than TE. In Fig. \ref{diff_timewidth} (a), the raw data for $T_p =$ 14 ns with $\Omega_c= 17.37 \Gamma$ are shown. Due to the non-optimized $\Omega_c$, the slowed and retrieved pulses have a significant broadening which leads to a reduced WL. In Fig. \ref{diff_timewidth} (b), we intentionally show an example with a significant pulse distortion or even splitting for the slowed pulse, in which $B < \Delta \omega_{EIT}$ is not satisfied. It is not surprised that both the efficiency and WL of the retrieved pulse are low under such a situation, which are 31.8\% and 64.7 \%, respectively.  

\subsection{Storage time}
\begin{figure}[t]
\includegraphics[width=8.8cm,viewport=240 30 760 310,clip]{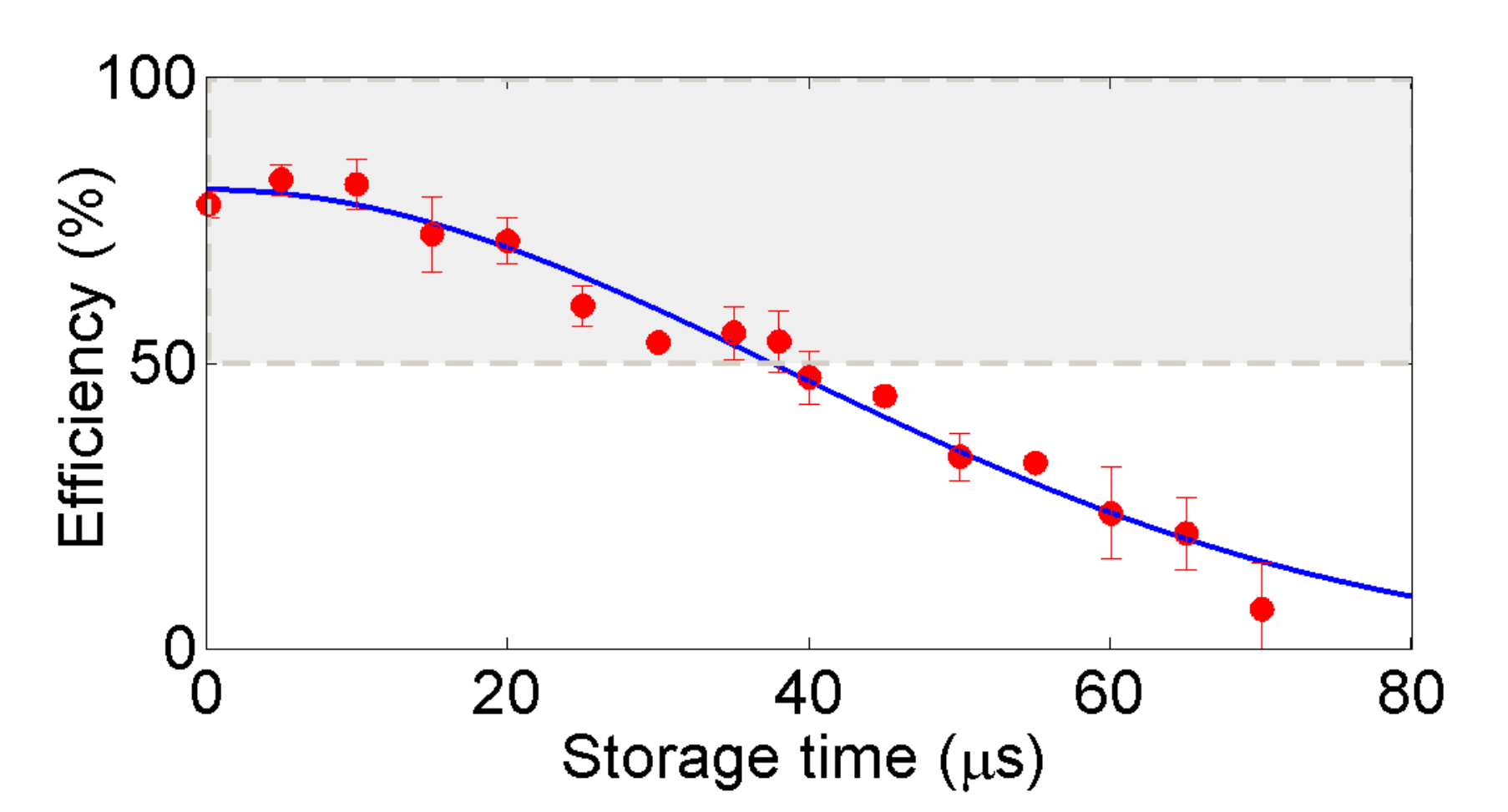}
\caption{ Efficiency versus the storage time. The blue solid line is a fitting curve to the data with the fit function: $A*\text{Exp}[-(t/\tau)^2]$, where the fitting parameters $A=80.3\%$ and $\tau=54.5 \mu$s. The grey dashed line represents the $50\%$ efficiency.}
\label{Storage}
\end{figure}

We then study the efficiency versus the storage time with $T_p=$ 30 ns for the input pulses. We have minimized the stray magnetic field by prolonging the storage time through three pairs of compensation coils. Fig. \ref{Storage} depicts an example of efficiency versus the storage time, ranging up to 70 $\mu$s. The data is fit to a curve of $A*\text{Exp}[-(t/\tau)^2]$ with fitting parameters $A=80.3 \%$ and $\tau= 54.5 \mu$s. The time-bandwidth product (TBP), defined as the ratio of the storage time at $50\%$ efficiency to the FWHM input pulse duration ($T_p$), is an important figure of merit in quantum memory application. The determined TBP based on Fig. \ref{Storage} is 1267, slightly higher than our previous work of 1200 with $T_p$=200 ns\cite{PhysRevLett.120.183602}. The storage time may still be limited by the residual magnetic field and the residual Doppler broadening due to the atomic motion and the finite angle ($\sim 1^0$) between the probe and control beams\cite{PanJW2009}.  

\section{Conclusion}
In summary, we explore the EIT-based storage towards broadband regime. The requirements for high-performance broadband EIT memory have been theoretically discussed. Large optical depth is necessary to reach the high-performance storage for short pulses, and waveform likeness becomes the limit when reaching the broadband regime. We experimentally demonstrate the broadband EIT memory with a storage efficiency of $\sim80 \%$ (WL 92.6 \%) for a 30-ns pulse and of $> 50\%$ (WL 79.7 \%) for a 14-ns pulse. The achieved time-bandwidth product is 1267. Our work clarify that it is possible to obtain a high-efficiency and a high-bandwidth for adiabatic EIT memory, provided one can achieve a high optical depth and a strong control intensity. 

\appendix
\section{Leakage-induced Loss}
\label{app: eta}
In this appendix, we consider the relation between the choice of the parameter $\eta$ and the optimized operational efficiency $F$ due to the pulse leakage during storage process for a given transmission efficiency (TE) of the slow light pulse, as plotted in Fig. \ref{Fig2} (b). We assume the EIT storage is in the regime that the dispersion up to O($\omega^2$) is a good approximation. As mentioned in the theoretical section, the slow light pulse is broadened by a factor of $\beta$ (Eq. \ref{beta}) and the relation $\text{TE}=\frac{1}{\beta}$ is satisfied in the ideal limit of $\gamma_{gs}= 0$. The overall operating efficiency $F$ is the product of two terms, $F_i$ and $F_o$, due to the leakage of the front and real tail of the pulse, respectively, which read\cite{PhysRevLett.120.183602},
\begin{equation}
\begin{aligned}
    F_i =  \frac{1}{2}(1 + erf(2 \sqrt{ln2} \kappa))\\
    F_o =\frac{1}{2}(1 + erf(2 \sqrt{ln2} (\eta - \kappa)/ \beta  ))\\
    F =& F_i \times F_o
\end{aligned}
\end{equation}
Here, $\kappa$ denotes the ratio of the turned-off time of the control field ($T_c$) to the temporal width of the input pulse, i.e. $\kappa \equiv \frac{T_c}{T_p}$. 

To search for the optimal $\kappa$ that minimizes the leakage-induced loss, we take the derivative of $F$ with respect to $\kappa$,
\begin{equation}
\begin{aligned}
    &\partial_\kappa F = 0\\
    &e^{-(2 \sqrt{ln2} \kappa)^2} (\frac{1}{2}(1 + erf(2  \sqrt{ln2} (\eta - \kappa) /\beta  )))\\
   & - \frac{1}{\beta} e^{-(2 \sqrt{ln2} (\eta - \kappa)/\beta)^2} (\frac{1}{2}(1 + erf(2 \sqrt{ln2} \kappa))) = 0
    \end{aligned}
\end{equation}
If the broadening is not too severe such that $\beta \simeq 1$, then when $\kappa = \frac{1}{\beta}(\eta - \kappa)$ the derivative is approximately zero. This assumption is valid if TE is not much less than unity (see Sec.\ref{sec:pulse}). Under such an approximation, the optimized $\kappa$ is,
\begin{equation}
    \kappa = \frac{\eta}{1 +\beta} = \frac{\eta}{1 +\frac{1}{\text{TE}}},
\end{equation}
and
\begin{equation}
    F_i = F_o.
\end{equation}
The operating efficiency is then, 
\begin{equation}
F = F_i F_o = F_i^2,    
\end{equation}
with
\begin{equation}
\begin{aligned}
    F_i =F_o= \frac{1}{2}(1 + erf(2 \sqrt{ln2} \kappa).
    \end{aligned}
\end{equation}

\bibliography{main}

\end{document}